%% file: paper.tex
\pdfoutput=1

\documentclass[]{bytedance_seed}

\usepackage[toc,page,header]{appendix}

\usepackage{minitoc}

\usepackage{microtype}
\usepackage{graphicx}
\usepackage{booktabs} 
\usepackage{hyperref}
\usepackage{multirow}
\usepackage{algorithmic}
\usepackage{amsmath}
\usepackage{amsfonts}
\usepackage{xspace}

\def\ie{\textit{i.e.}\xspace}

\newcommand{\ours}{\textsc{TileLink}}

\title{ \ours{}: Generating Efficient Compute-Communication Overlapping Kernels using Tile-Centric Primitives  }

\author[1,*,\dagger]{Size Zheng}
\author[1,*]{Jin Fang}
\author[1]{Xuegui Zheng}
\author[1]{Qi Hou}
\author[1]{Wenlei Bao}
\author[1]{Ningxin Zheng}
\author[1]{Ziheng Jiang}
\author[1]{Dongyang Wang}
\author[1]{Jianxi Ye}
\author[1]{Haibin Lin}
\author[1]{Li-Wen Chang}
\author[1, \dagger]{Xin Liu}

\affiliation[1]{ByteDance Seed}

\contribution[*]{Equal contribution}
\contribution[\dagger]{Corresponding authors}

\abstract{

\input{contents/abstract}
}

\date{\today}
\correspondence{Xin Liu at \email{liuxin.ai@bytedance.com}, Size Zheng at \email{zheng.size@bytedance.com}}

\checkdata[Project Page]{\url{https://github.com/ByteDance-Seed/Triton-distributed}}

\begin{document}
\maketitle


\input{contents/introduction}
\input{contents/background}

\input{contents/overview}
\input{contents/frontend}
\input{contents/backend}

\input{contents/optimization}

\input{contents/compiler}
\input{contents/evaluation}
\input{contents/related-work}
\input{contents/conclusion}

\clearpage

\bibliographystyle{plainnat}
\bibliography{refs}

\clearpage



\end{document}

%% file: contents/abstract.tex
Large deep learning models have achieved state-of-the-art performance in a wide range of tasks. These models often necessitate distributed systems for efficient training and inference. The fundamental building blocks for distributed model execution are intra-layer parallel operators.
The most effective approach to enhancing the performance of intra-layer parallel operators involves overlapping computation with communication. The overlapping can be achieved through either operator decomposition or kernel fusion. 
While decomposing operators is straightforward to implement, it often results in suboptimal performance. On the other hand, fusing communication kernels with compute kernels demands significant expertise and is error-prone.

In this paper, we propose \ours{} to enable efficient compilation and generation of overlapped compute-communication kernels. \ours{} is composed of frontend and backend.
In the frontend, \ours{} decouples the design space of communication and computation, linking these two parts via tile-centric primitives.
In the backend, \ours{} translates these primitives into low-level communication instructions, integrating the communication and computation components to achieve overlapped execution.
In experiments, \ours{} achieves from $1.17\times$  to $20.76\times$ speedup to non-overlapping baseline and achieves performance comparable to state-of-the-art overlapping libraries on GPUs.


%% file: contents/introduction.tex
\section{Introduction}

Large deep learning models keep growing in both model size and performance. These models have achieved state-of-the-art results in a wide range of domains including natural language processing~\cite{gpt4, gemma2, llama3, deepseek-v2}, vision processing~\cite{clip, minigpt4, qwen-vl, deepseek-vl}, and reasoning~\cite{o1, deepseek-math}. The substantial sizes of these models, coupled with their immense computational demands, necessitate parallel execution across distributed systems. Various parallel methods have been proposed to accelerate distributed processing by exploiting both intra-layer and inter-layer parallelism~\cite{zero, megatron-lm, gpipe, deepspeed}.

Since intra-layer parallelism forms the foundation of parallel computing, a significant body of work has focused on exploring it~\cite{megatron-lm, te, ringattn}. While parallel execution enhances overall performance, communication between devices still incurs significant overhead, limiting further improvements in computational efficiency~\cite{centauri, flux}. Previous work~\cite{flux} indicates that communication overhead constitutes approximately $10\%$ to $50\%$ of the total execution overhead even in machines equipped with high-speed inter-device links.

Overlapping communication with computation is an effective strategy for enhancing computational efficiency. The core idea is to map communication and computation to distinct hardware units, allowing them to operate concurrently. To handle data dependency between communication and computation operators, synchronization or barriers are inserted into the loop of data transfer and computation. Previous work on overlapping communication with computation mainly uses two techniques: operator decomposition and kernel fusion.

Operator decomposition~\cite{te, dist-enisum, centauri} involves breaking down both communication and computation kernels into smaller, homogeneous kernels. The data dependencies are then distributed across multiple communication-computation kernel pairs. The smaller kernels, once split, can be dispatched to different streams, allowing communication and computation kernels to operate on separate data shards simultaneously. 
Operator decomposition can be easily implemented on modern deep learning frameworks such as PyTorch~\cite{pytorch} or TensorFlow~\cite{tensorflow}, enabling systematic exploration of the entire design space for communication-computation overlap, including model-level, layer-level, and operation-level as pointed out in previous work~\cite{centauri}. 
However, synchronization between these decomposed kernels necessitates host intervention, introducing non-negligible overhead at runtime. Furthermore, the performance of decomposed kernels may be degraded due to low cache utilization and resource quantization inefficiency.

On the other hand, the kernel fusion method~\cite{coconet, flux} combines communication and computation kernels into one fused kernel to overcome the above disadvantages. Within fused kernels, communication is mapped to either DMA (Direct Memory Access) engines or processing cores (e.g., streaming multiprocessors on a GPU), while computation is executed simultaneously on other processing cores. Data dependencies are managed using on-device barriers, and processing cores responsible for data transfer communicate with computation cores through atomic or communication instructions. This method is efficient in terms of performance but often requires high-level hardware expertise to implement efficient kernels, and it struggles to keep pace with rapid algorithm development.

To address the challenges inherent in existing approaches, we propose \ours{}, a framework designed to enhance the development efficiency of overlapping kernels through compilation. \ours{} consists of two main components: a frontend and a backend.
In the frontend, \ours{} decouples the design space of communication and computation kernels, enabling each to utilize distinct optimization strategies and tiling methods. To allow the communication and computation kernels to operate with different tile sizes, it relies on tailored barrier controls to maintain producer-consumer dependencies, ensuring correct and efficient execution.
Typically, this fusion is achieved by directly programming in assembly. To automate the fusion of communication and computation kernels without requiring low-level assembly code, \ours{} offers a set of tile-centric primitives. These primitives provide abstract semantics for signaling and data communication between devices, while concealing low-level details such as pointer management and barrier control.

In the backend, \ours{} compiles tile-centric primitives into low-level hardware instructions, integrating them with the communication and computation kernels. To ensure correct data exchange and barrier manipulation operations, \ours{} employs a tile-centric mapping strategy, which includes shape mapping, rank mapping, and channel mapping.
This tile-centric mapping can be either static or dynamic. Static mapping uses affine transformations at compile time to map tile IDs to shape ranges, rank IDs, and communication channels. In contrast, dynamic mapping computes these mappings on-the-fly at runtime, allowing greater flexibility.
To show the flexibility and generality of \ours{}, we implement a broad range of overlapped workloads using \ours{}, including self-attention, MLP (multilayer perceptron), and MoE (mixture of experts).
In addition to programming efficiency, \ours{} also achieves high performance on GPUs.
Evaluation on 8$\times$H800 GPUs shows that \ours{} can achieve from $1.17\times$ to $20.76\times$ speedups over non-overlapping baselines, achieving comparable or better performance to overlapping libraries, such as FLUX~\cite{flux} and RingAttention~\cite{ringattn}. For end-to-end evaluation, we test eight different language models on 8$\times$H800 GPUs and the average speedup of \ours{} is $1.32\times$ compared to PyTorch. We also benchmark \ours{} on two nodes of 8$\times$H800 (totally 16 GPUs) and the average speedup to PyTorch is $1.29\times$.

%% file: contents/background.tex
\section{Background}

\begin{figure}[!t]
\centering
\includegraphics[width=0.49\textwidth]{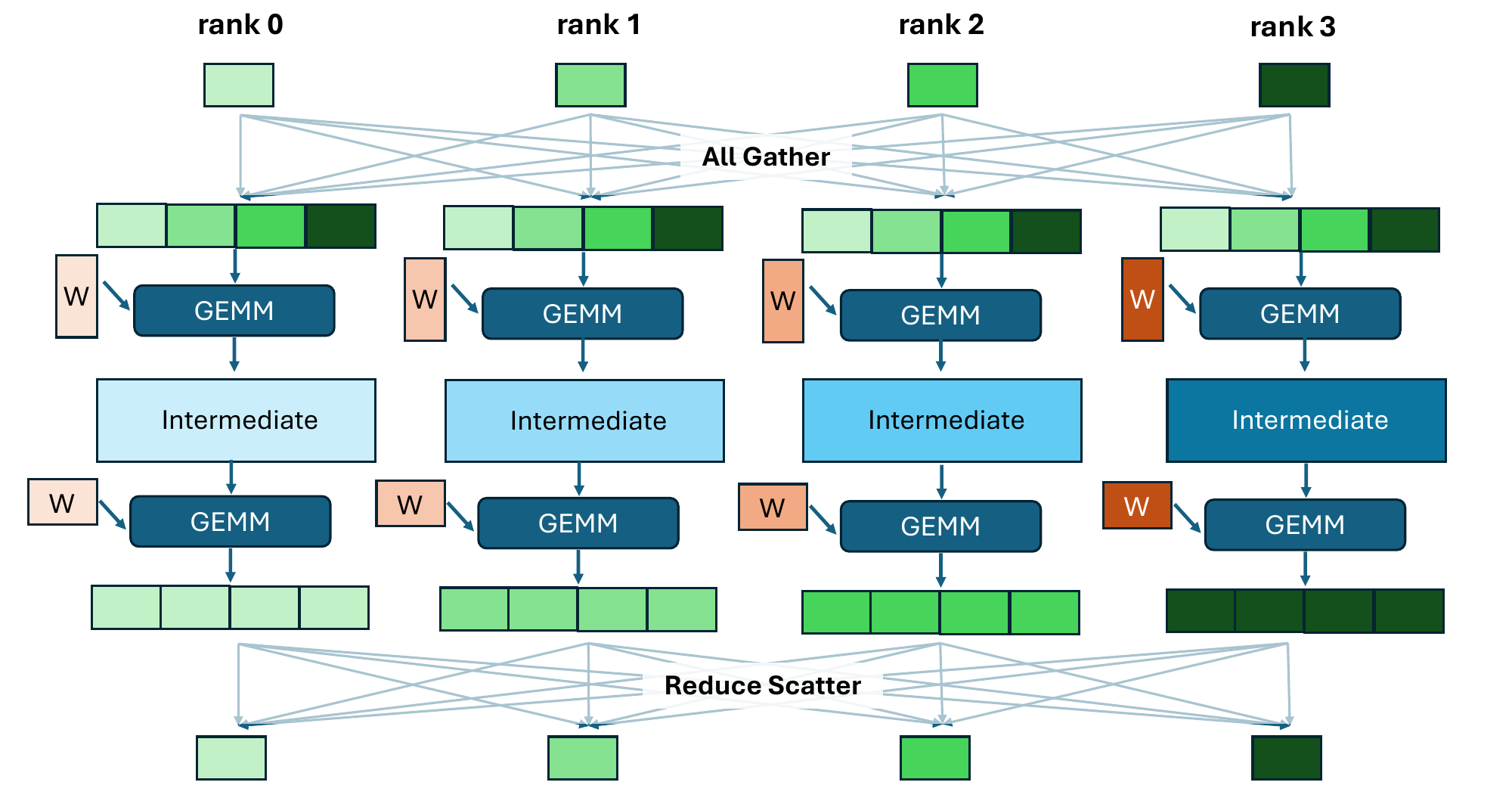}
\DeclareGraphicsExtensions.
\caption{Intra-layer parallel FFN example.
}
\label{fig:ffn}
\end{figure}


\subsection{Communication and Intra-Layer Parallelism}

\textbf{Operator-Centric Communication Primitives:} Collective communications are frequently used in parallel execution of large models. Existing libraries~\cite{nccl} and frameworks~\cite{pytorch, tensorflow} employ operator-level primitives for common communication patterns, such as AllReduce, ReduceScatter, AllGather, and All2All. These primitives need to perform system synchronization before and after data transfer to follow the SPMD (Single Program, Multiple Data) programming model and integrate seamlessly with other operators. However, coarse-grained synchronization can cause computational units to be idle during communication, thus reducing computational efficiency. We call these communication primitives \textit{operator-centric primitives}.

\textbf{Intra-Layer Parallelism with Operator-Centric Primitives:}
For large models (\ie, Transformer-based models), intra-layer parallelism is primarily applied to two components: the attention part and the FFN (feed-forward network) part, which is composed of MLP (multilayer perceptron) layers or MoE (mixture of experts) layers. For the attention part, the context (key and value) is sharded across devices. Before computation, these context shards are gathered to form a complete context for self-attention. This parallel algorithm is referred to as sequence-parallel~\cite{megatron-lm, ringattn}.

For the FFN part, the weights of the two layers in the MLP or MoE are sharded across devices. First, input data is gathered from different ranks, followed by local computation using the corresponding weight shards. Finally, the partial results are reduced and scattered to the appropriate ranks. This algorithm, commonly used in previous work~\cite{megatron-lm, coconet, dist-enisum}, is depicted in Figure~\ref{fig:ffn} and is referred to as tensor-parallel FFN. Using existing communication libraries and frameworks, tensor-parallel FFN is expressed as AllGather + GEMM (or GroupGEMM) followed by GEMM (or GroupGEMM) + ReduceScatter.

\subsection{Communication and Computation Overlapping}

Overlapping communication and computation has been extensively explored in prior studies~\cite{centauri, coconet, dist-enisum, flux}. Centauri~\cite{centauri} introduces a comprehensive three-level design space encompassing model-level, layer-level, and operation-level overlapping. Intra-layer parallelism serves as the foundational element across these levels of overlapping. \ours{} focuses on intra-layer overlapping; generalizing \ours{}'s techniques to inter-layer or model-level overlapping is feasible, but beyond the scope of this paper.
For intra-layer parallelism, there are two main ways to achieve overlapping: operator decomposition and kernel fusion. As shown in Table~\ref{table:related-work}, we summarize the features of representative studies and \ours{}.

\textbf{Operator Decomposition:} 
This approach splits the original communication and computation operators into smaller, fine-grained units. These smaller operators enable more precise synchronization control, allowing communication operators to execute in parallel with computation operators that do not have data dependencies.
Operator decomposition is advantageous due to its straightforward implementation and compatibility with existing libraries and frameworks. However, using smaller operators can lead to inefficiencies, including low L2 cache utilization \cite{ark} and resource quantization inefficiency \cite{streamk}. Additionally, synchronization between kernels requires host intervention, introducing non-negligible overhead. Representative works employing operator decomposition include Dist-Einsum \cite{dist-enisum}, Asynchronous Tensor Parallel PyTorch \cite{pytorch}, and Centauri \cite{centauri}.

\begin{table}[t]
  \centering
  \footnotesize
  \caption{Feature comparison of \ours{} and previous work.}
\begin{small}
  \begin{tabular}{c|c|c|c}
  \hline
    \textbf{Name} & \textbf{Compile} & \textbf{Method} & \textbf{Primitive}\\ 
    
  \hline
    \textbf{CoCoNet} & Yes & Fusion & No\\

  \hline
    \textbf{Dist-Einsum} & Yes & Decompose  & operator-centric\\


  \hline
    \textbf{Centauri} & No & Decompose  & operator-centric\\
    
  \hline
    \textbf{FLUX} & No & Fusion & No\\

  \hline
    \textbf{Async-Torch} & No & Decompose & operator-centric\\

  \hline
    \textbf{\ours{}} & Yes & Fusion & tile-centric\\
  \hline
  \end{tabular}
  \label{table:related-work}
\end{small}
\end{table}

\begin{table}[t]
  \centering
  \footnotesize
  \caption{Motivational example.}
\begin{small}
  \begin{tabular}{c|c|c}
  \hline
    \multicolumn{3}{c}{{Configurations of TP MLP}} \\
  
  \hline
    \textbf{batch$\times$sequence length} & \textbf{hidden dim} & \textbf{intermediate size}\\ 

  \hline
    8192 & 4096 & 11008 \\

  \hline
    \multicolumn{3}{c}{{Performance of Different Overlapping Techniques}} \\

  \hline
    \multirow{2}{*}{\textbf{Method}} & \multicolumn{2}{c}{\textbf{Performance}}\\

  \cline{2-3}
     & \textbf{AG+GEMM} & \textbf{GEMM+RS}\\

  \hline
    Non-Overlap & 0.676 ms & 0.541 ms \\

  \hline
    Decomposition & 1.301 ms & 1.443 ms  \\

  \hline
    Fusion (FLUX) & 0.504 ms & 0.610 ms \\

  \hline
    \ours{} (ours) & 0.505 ms & 0.504 ms \\
    
  \hline
  \noalign{\smallskip}
  \hline
   \multirow{2}{*}{\textbf{Lines of Code}} & FLUX & \ours{} (ours)\\
    & $\approx$ 2,000 .cu & $\approx$ 200 .py\\

  \hline
  \end{tabular}
  \label{table:mlp-motivation}
\end{small}
\end{table}

\textbf{Kernel Fusion:} This approach fuses communication and computation kernels. Typically, the fused kernel allocates part of the processing cores to communication tasks and the remaining cores to computation tasks. Cores assigned to different tasks use on-device barriers to communicate execution states. 
The fused kernel eliminates the need for host intervention during synchronization, improves cache utilization, and mitigates resource quantization inefficiency, potentially achieving better performance than the operator decomposition method. However, developing fused kernels on modern accelerators, such as GPUs, presents challenges. On one hand, low-level control over barriers and hardware-related communication instructions demands a high level of expertise. On the other hand, improper fusion design may lead to performance degradation due to resource conflicts between communication and computation cores. 
Consequently, only a few highly optimized libraries~\cite{amd-fused, flux} or domain-specific compilers~\cite{coconet} support the kernel fusion method. 

\subsection{Code Generation Compilers}

With the rapid advancement of code generation compilers~\cite{halide, tvm, triton}, generating high-performance code for attention or FFN has become practical.
Although previous overlapping compilers such as CoCoNet~\cite{coconet} and Dist-Einsum~\cite{dist-enisum} can generate overlapped kernels, they are restricted to fixed overlapping patterns without programming flexibility at the operator level.
In contrast, \ours{} uses tile-centric primitives and enables efficient compilation for a variety of workloads.

\subsection{Motivational Example}

To illustrate the benefits of \ours{}, we use a tensor-parallel MLP layer as a motivational example.
The input shape of the MLP layer, detailed in Table~\ref{table:mlp-motivation}, corresponds to the configuration used in the LLaMA-7B model. This MLP layer is implemented as AllGather + GEMM (AG + GEMM) followed by GEMM + ReduceScatter (GEMM + RS), as depicted in Figure~\ref{fig:ffn}. We compare the performance of different techniques for these two parts in Table~\ref{table:mlp-motivation}. \textit{Non-Overlap} is to use cuBLAS~\cite{cublas} and NCCL~\cite{nccl} with no overlapping. \textit{Decomposition} uses the operator decomposition technique, with performance results taken from Async-TP PyTorch~\cite{torchtitan}. 
\textit{Fusion} refers to the kernel fusion technique, measured using the open-source library FLUX~\cite{flux}.

On one hand, we compare the performance achieved by different techniques. As shown in the Table~\ref{table:mlp-motivation}, the decomposition technique delivers the lowest performance, while the fusion technique achieves the best results for AG + GEMM.
\ours{} achieves the best performance for GEMM + RS and comes very close to FLUX for AG + GEMM (about $99\%$). These findings demonstrate that \ours{} is capable of delivering performance that is comparable to or better than previous approaches.
On the other hand, we compare the lines of code required by FLUX and \ours{}. FLUX involves approximately 2,000 lines of CUDA code, whereas \ours{} achieves similar performance with only around 200 lines of Python code, resulting in a roughly $10\times$ improvement in programming efficiency. This motivational example highlights the significant advantages of \ours{}.

%% file: contents/overview.tex



%% file: contents/frontend.tex
\section{Frontend Primitives}

In this Section, we explain the frontend of \ours{}. We first explain the decoupled design space. Then, we present \ours{}'s tile-centric primitives.

\begin{table*}
  \hypersetup{hidelinks}
  \centering
  \footnotesize
  \caption{Tile-centric primitives in \ours{}}
  \begin{tabular}{c|c|c}
  \hline
    \textbf{Name} & \textbf{Usage} & \textbf{Explanation} \\ 
    







  \hline
    \multirow{3}{*}{\textit{producer\_tile\_notify}} & \multirow{3}{*}{\textit{producer\_tile\_nofity(tile\_id, mode)}} & Mark producer tile done and notify its consumer tile, \\
    
    & & consumer tile is marked ready when all the producer tiles \\
    & & it depends on are done \\

  \hline
    \multirow{2}{*}{\textit{consumer\_tile\_wait}} & \multirow{2}{*}{\textit{consumer\_tile\_wait(tile\_id)}} & Consumer tile is blocked until all its \\
    & & dependent producer tiles done\\

  \hline
    \multirow{2}{*}{\textit{peer\_tile\_notify}} & \multirow{2}{*}{\textit{peer\_tile\_notify(tile\_id, rank)}} & Mark current tile done and notify  \\
    & & its peer tiles in another rank\\

  \hline
    \multirow{2}{*}{\textit{peer\_tile\_wait}} & \multirow{2}{*}{\textit{peer\_tile\_wait(tile\_id, rank)}} & Block current tile until its peer tile  \\
    & & in another rank is done\\

  \hline
    \textit{rank\_notify} & \textit{rank\_notify(tile\_id, rank)} & Tell another rank that data at tile\_id is ready \\

  \hline
    \textit{rank\_wait} & \textit{rank\_wait(rank)} & Block current rank until another rank is done  \\

  \hline
    \multirow{2}{*}{\textit{tile\_push\_data}} & \multirow{2}{*}{\textit{tile\_push\_data(tensors, tile\_id, data)}} & Send a tile of data to one (p2p)  \\
    & & or all the (broadcast) remote tensors\\

  \hline
    \multirow{2}{*}{\textit{tile\_pull\_data}} & \multirow{2}{*}{\textit{data = tile\_pull\_data(tensors, tile\_id)}} & Load one (p2p) or all the (broadcast) \\
    & & tiles of data from remote tensors\\


  \hline
    \textit{rank\_copy\_data} & \textit{rank\_copy\_data(src, dst)} & Copy data from src rank to dst rank\\

  \hline
  \end{tabular}
  \label{table:primitives}
\end{table*}

\subsection{Decoupled Design Space}

There are two ways to design compute-communication fusion kernels. One is to tightly couple the optimization choices of the two parts, including tile size, tile order, and resource mapping, while the other is to decouple computation and communication kernel design. \ours{} adopts the latter one because the decoupled design space enables more flexibility in kernel design and could result in better performance.

\begin{figure}[!t]
\centering
\includegraphics[width=0.49\textwidth]{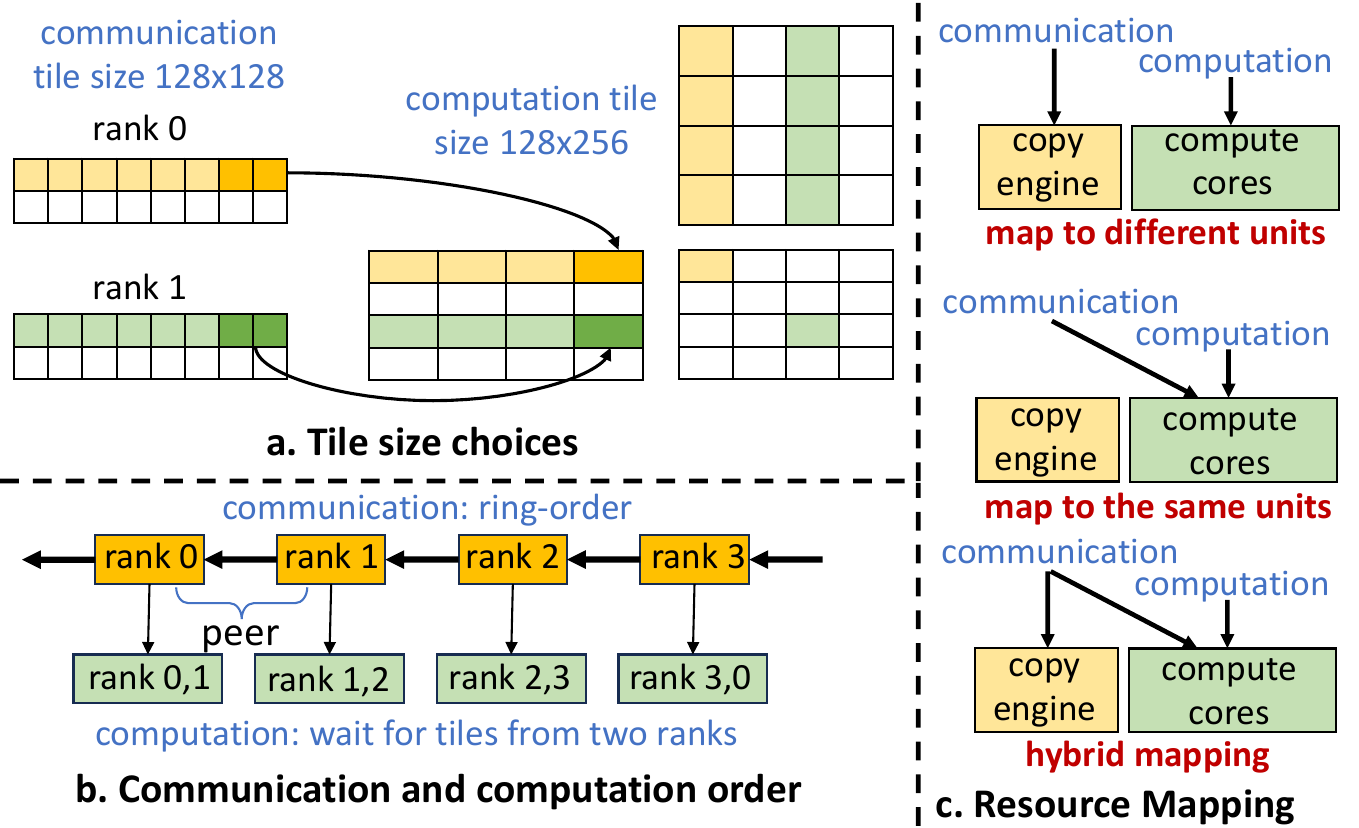}
\DeclareGraphicsExtensions.
\caption{Examples of the three design sub-spaces of communication and computation.}
\label{fig:design-space}
\end{figure}

We divide the decoupled design space into three subspaces: tile size, tile order, and resource mapping. 
For each of these subspaces, the communication and computation components can make independent choices to optimize their performance.
In the tile size subspace, the communication and computation components can choose different tile sizes. For example, as illustrated in Figure~\ref{fig:design-space}a, the communication part transfers a tile of $128\times 128$ at a time, while the computation part consumes a tile of size $128\times 256$ at a time. 
This differentiation in tile size helps each component achieve optimal performance by aligning with the number of processing cores it uses.
For instance, given an AllGather + GEMM problem with the tensor size of $M\times N\times K$, where the AllGather part binds dimension $M,K$ to processing cores, and the GEMM part binds dimension $M,N$. 
If the communication component uses more cores, a smaller tile size will be beneficial, because all core resources can be fully utilized; conversely, if it uses fewer cores, a larger tile size will be more effective.

\begin{figure}[!t]
\centering
\includegraphics[width=0.48\textwidth]{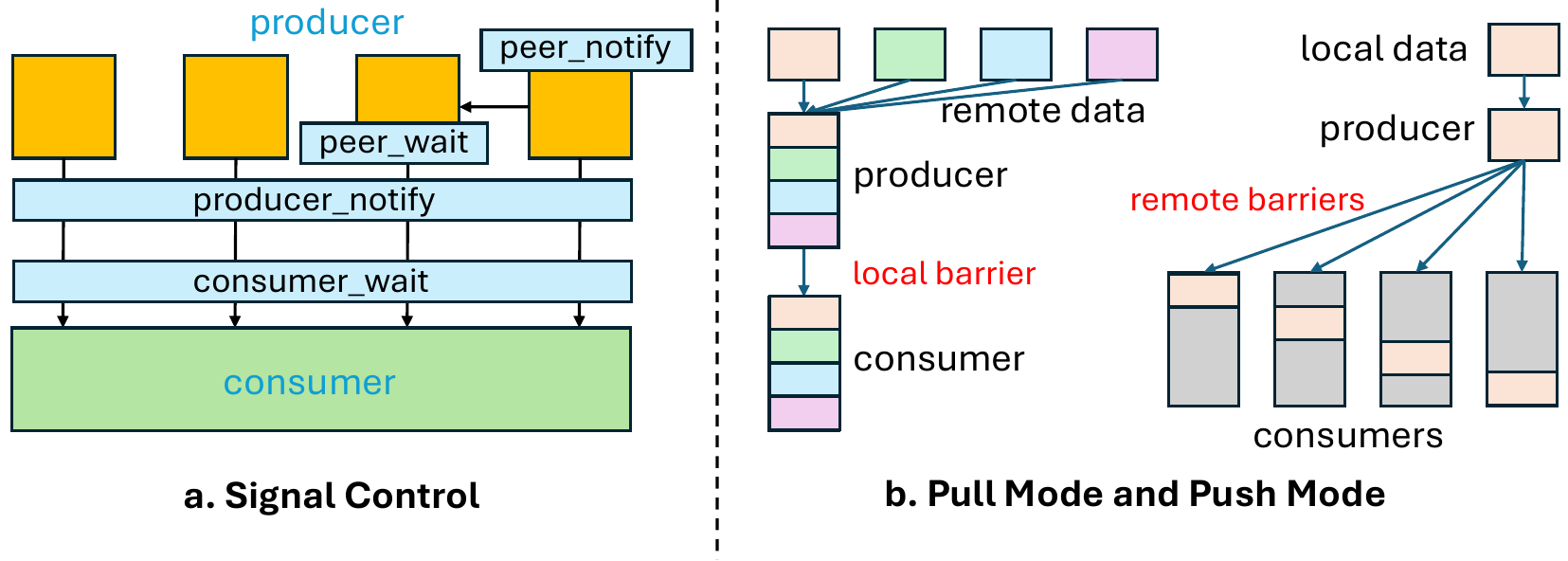}
\DeclareGraphicsExtensions.
\caption{Tile-centric primitives support different signal control and data transfer directions.}
\label{fig:tile-relation-view}
\end{figure}

In the tile order subspace, the communication component may utilize a different tile order compared to the computation component. For instance, communication can adopt various data transfer orders, such as ring order, full-mesh all-to-all order, or other patterns, while the computation component can begin processing data tiles from any rank.
There is a trade-off associated with the choice of tile order. If the computation component waits for data tiles from multiple ranks, it may achieve better cache efficiency when operating on larger chunks of data; however, this approach may result in longer wait time. Conversely, if the computation component only waits for data tiles from a single rank, it can start computations earlier, but this may lead to lower overall computation efficiency.
Figure~\ref{fig:design-space}b shows an example, where communication uses ring order and computation waits for data from two ranks at each iteration of computation.

In the resource binding subspace, the communication and computation components can be mapped to either different units or the same unit, as illustrated in Figure~\ref{fig:design-space}c.
If the communication component utilizes the copy engine (DMA), it avoids resource conflicts with the computation component. However, this approach involves host interference, which introduces additional overhead. On the other hand, if the communication component employs compute cores for data copying, it may lead to resource conflicts with the computation component but eliminates host overhead. This strategy is particularly suitable in scenarios where the computation component cannot fully utilize all available processing cores.

\subsection{Tile-Centric Primitives}
Decoupling the design space of communication and computation introduces synchronization challenges. Since the two components utilize different tile sizes, tile orders, and resource mappings, synchronizing them necessitates complex low-level programming with communication instructions. For example, on GPUs, instructions such as \textit{ld.global.acquire} and \textit{red.release} are required.
However, the programming model for these instructions does not align with that of code generation compilers \cite{tvm, triton}, as existing compilers lack support for a memory consistency model.

To address this issue, \ours{} offers a set of tile-centric primitives. These primitives introduce memory consistency semantics and adhere to the tile-level abstraction utilized in the compiler, distinguishing them from the operator-centric primitives provided by previous frameworks \cite{pytorch, tensorflow} and libraries \cite{nccl}.
The primitives of \ours{} are summarized in Table~\ref{table:primitives}. They are categorized into two groups: signal primitives and data primitives. Each group contains both device-side primitives and host-side primitives.

\subsubsection{Signal Primitives}

Signal primitives are designed to manage barriers between communication and computation. They include \textit{producer(peer)\_tile\_notify}, \textit{consumer(peer)\_tile\_wait}, and \textit{rank\_notify(wait)}.
For device-side primitives, the \textit{producer\_tile\_notify} and \textit{consumer\_tile\_wait} primitives are applied to producer-consumer relationships, such as those between the tiles of AllGather and GEMM. The \textit{peer\_tile\_notify} and \textit{peer\_tile\_wait} primitives are primarily used for tiles of the same operator across different ranks, enabling users to construct various tile orders.
For host-side primitives, the \textit{rank\_notify(wait)} primitive is used to manage barriers between the copy engine and compute cores. When communication is mapped to the copy engine, these primitives facilitate the control of tile orders between communication and computation.
Figure~\ref{fig:tile-relation-view}a shows the signal control between communication and computation parts.

Notify primitives require either \textit{mode} argument or \textit{rank} argument to clarity which remote ranks to notify.
\ours{} provides two choices for \textit{mode} argument: \textit{p2p} and \textit{broadcast}. \textit{p2p} means that only one target rank will be notified, which is calculated by the offset of the given \textit{tile\_id} in the global tensor view; \textit{broadcast} means that all the ranks will be notified.

\textbf{Memory Consistency:} 
In parallel executions, memory operations performed by different processes/threads may become visible to others in a non-uniform order. The memory consistency model specifies constraints to prevent contradictions in the observed order of operations across processes/threads.
The signal primitives provide strict memory consistency semantics. The notify primitives carry release semantics, ensuring that any memory access occurring before \textit{producer(peer)\_tile\_notify} and \textit{rank\_notify} cannot be executed after these notify primitives. Conversely, the wait primitives carry acquire semantics, ensuring that any memory access following \textit{consumer(peer)\_tile\_wait} and \textit{rank\_wait} cannot be executed before these wait primitives.
This strict memory consistency must also be taken into account during backend compilation, which will be discussed later.

\subsubsection{Data Primitives}
Data primitives facilitate data transfer and include \textit{tile\_push(pull)\_data} and \textit{rank\_copy\_data} primitives. These primitives control the resource mapping and tile sizes of the transferred data.
The device-side \textit{tile\_push(pull)\_data} primitive maps communication to processing cores, while the  host-side \textit{rank\_copy\_data} primitive maps communication to the copy engine. There are two modes for data transfer—pull and push—each suited for different synchronization methods. In the pull mode, the producer reads data from all other ranks and notifies its consumer using local barriers. In contrast, the push mode allows the producer to write local data to all other ranks while notifying its remote consumers of the data's arrival.
Figure~\ref{fig:tile-relation-view}b illustrates the differences between the two modes. The choice between pull and push modes may impact performance (as pointed out in FLUX~\cite{flux}), depending on factors such as data shapes, tiling strategies, and available hardware resources. Notably, the \textit{rank\_copy\_data} primitive supports both modes through peer-to-peer copying, with the data transfer direction indicated by the order of the source and destination pointers.

%% file: contents/backend.tex
\section{Backend Mapping}

The backend of \ours{} handles the compilation of both communication and computation components into low-level device codes. To enable code generation for distributed systems, \ours{} employs a tile-centric mapping technique that links parts of the communication and computation.
In this section, we first explain the tile-centric mapping approach and the compilation process used by \ours{}. Next, we describe how \ours{} ensures memory consistency. Finally, we briefly summarize additional compilation techniques applied for single device.

\subsection{Tile-Centric Mapping}

\ours{} uses a tile-centric mapping approach to compile frontend primitives into low-level code. Tile-centric mapping consists of three components: shape mapping, rank mapping, and channel mapping.
Shape mapping associates each \textit{tile\_id} with a specific tensor shape slice.
Rank mapping links each \textit{tile\_id} to a device rank.
Channel mapping assigns each \textit{tile\_id} to a communication barrier.
We use $f_S, f_R, f_C$ to represent these three mappings, respectively. Depending on the workload type, different mapping functions should be used. We classify the different mappings into two groups: static mapping and dynamic mapping.

Static mapping refers to mappings that can be decided at compile time.
Static mapping is commonly used when data sharding strategy is fixed such as tensor-parallel MLP and sequence-parallel self-attention.
We use affine operations to handle static mapping ($f_S, f_R, f_C$ are affine).
For example, for AllGather (pull mode) + GEMM (problem size $M\times N\times K$) on $R$ ranks with $C$ channels per rank (each rank corresponds to $C$ barriers), the producer AllGather uses tile size $Tm_p \times Tn_p$, and the input tensor is sharded along $M$ dimension. Given producer tile $tile\_id_p$, the shape range, source rank, and channel can be computed as follows:
%
\begin{equation*}
\begin{small}
\begin{aligned}
& M\_per\_rank = \lceil \frac{M}{R} \rceil, \ \ \ M\_per\_channel = \lceil \frac{M}{R * C} \rceil,\\
& range_M = [tile\_id_p * Tm_p, tile\_id_p * Tm_p + Tm_p),\\
& src\_rank = \lfloor \frac{tile\_id_p}{\lfloor \frac{M\_per\_rank}{Tm_p} \rfloor} \rfloor,\ \ channel = \lfloor \frac{tile\_id_p}{\lfloor \frac{M\_per\_channel}{Tm_p} \rfloor} \rfloor.\\
\end{aligned}
\end{small}
\end{equation*}
Similarly, we can compute the mapping from consumer $tile\_id_c$ to shape range, rank, and channel.

Dynamic mapping refers to mappings computed at runtime, which are essential for workloads with dynamic data sharding requirements. For example, in the MoE data sharding strategy, dynamic routing determines the data distribution, and each tile may require tokens from any other rank. It is impossible to determine from which ranks to gather data or at which channel to wait for a barrier at compile time. Consequently, the mapping must be computed at runtime. To support dynamic mapping, \ours{} transforms these mappings into lookup tables, whose values can be filled at runtime, while the access operations to these lookup tables are determined at compile time. Formally, dynamic mapping is
\begin{equation*}
\begin{small}
\begin{aligned}
& range = [f_S\_low[tile\_id], f_S\_high[tild\_id]),\\
& rank = f_R[tile\_id], \ \ \ channel = f_C[tile\_id],
\end{aligned}
\end{small}
\end{equation*}
where $f_S\_low, f_S\_high, f_R$ and $f_C$ are lookup tables, the values of them will be filled at runtime by other dynamic logics (e.g., dynamic routing).

\subsection{Compilation for Memory Consistency}
In backend compilation, the frontend primitives with memory consistency semantics are compiled to corresponding device instructions such as \textit{ld.global.acquire} and \textit{red.release}. However, directly translating these primitives is not enough to guarantee memory consistency. For most computation kernels, multi-stage pipeline is applied to enhance load-compute balance and improve overall performance. Compiling original programs into multi-stage version requires operator reordering, during which some memory access operations may be reordered before or after \ours{} primitives unexpectedly. To address this issue, \ours{} enforces strict data dependencies between its primitives and their following load/store operations so that its primitives can be correctly reordered and unrolled by pipelining passes.

\subsection{Other Compilation Optimizations}

Apart from the aforementioned techniques, \ours{} also leverages strategies for single-device optimization to achieve high performance, which has been well-studied in previous work~\cite{tvm, triton}. The optimizations primarily include two aspects: memory optimization and pipeline optimization.
Memory optimization involves the automatic allocation of on-chip register buffers and shared memory buffers for computation. Data access to global buffers is coalesced, and the access pattern to shared memory is transformed to avoid bank conflicts.
Pipeline optimization involves rearranging data load/store operations and computations to form a multi-stage pipeline. Local data copies are mapped to dedicated asynchronous engines, such as the Tensor Memory Accelerator (TMA) of GPUs. Computation is mapped to high-performance units, such as the Tensor Core units of GPUs.

%% file: contents/optimization.tex
\section{Kernel Design with \ours{}}

\begin{figure}[!t]
\centering
\includegraphics[width=0.47\textwidth]{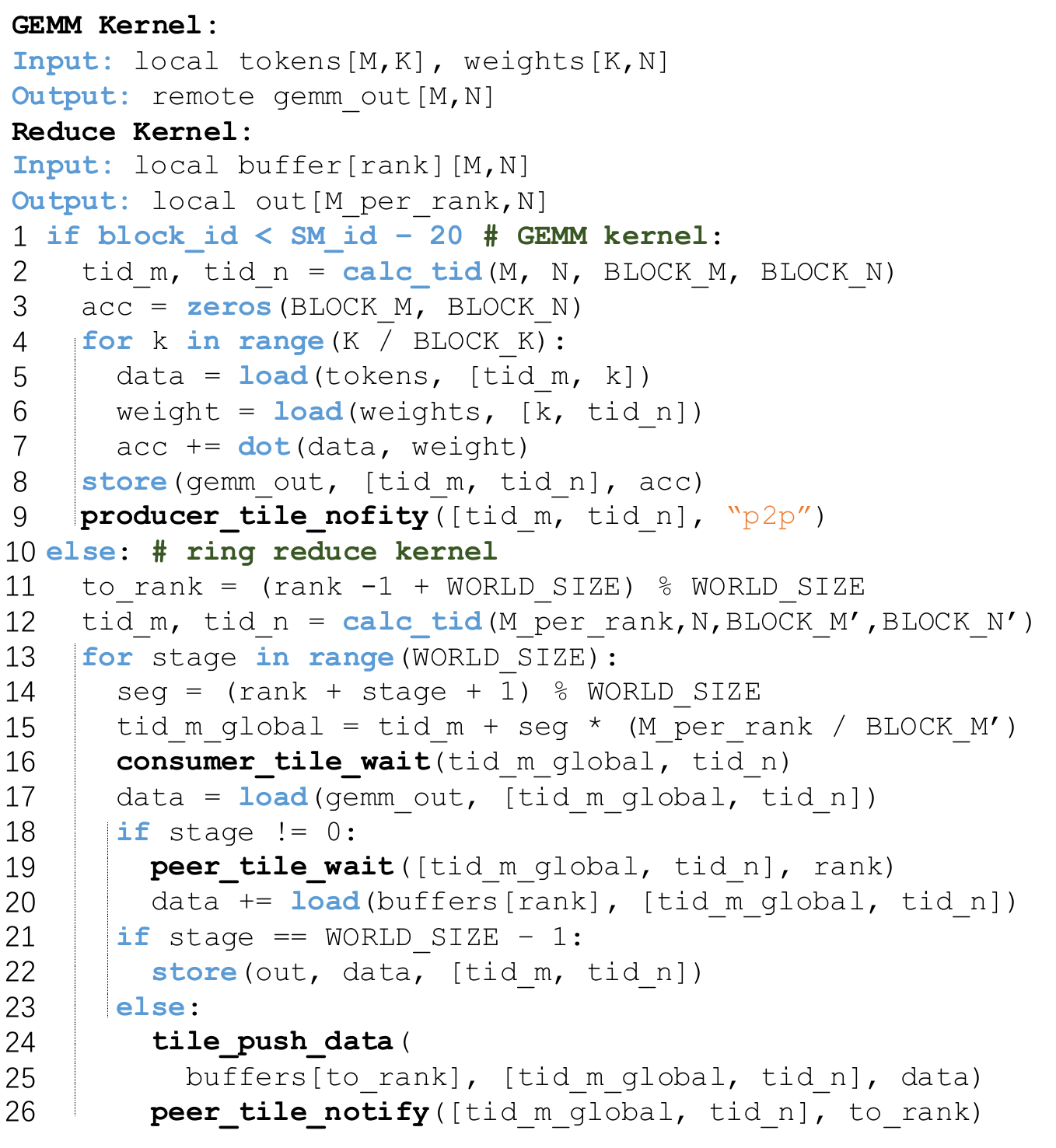}
\DeclareGraphicsExtensions.
\caption{GEMM+RS overlapping kernel using \ours{}.}
\label{fig:gemm_rs}
\end{figure}



To demonstrate the flexibility and generality of \ours{}, we present how to design overlapped kernels for GEMM + ring ReduceScatter, AllGather + MoE, and AllGather KV + self-attention. These three examples are representative because they utilize different tile orders (ring and all-to-all), different mappings (static and dynamic), and difference hardware resources (device and host).

Figure~\ref{fig:gemm_rs} shows the pseudo code for the GEMM + ring ReduceScatter kernel. Both computation and communication use SMs, we use 20 SMs for communication in this example (see line 1). The producer GEMM stores partial outputs in the local tensor and notifies its consumer using \textit{producer\_tile\_notify} (line 9). The consumer ReduceScatter waits for its producer at line 16. Once the data from the producer is ready, the consumer kernel performs a local reduction (line 20) and passes the partial results to its previous rank (line 24). Signal control between peer ranks is managed using the primitives \textit{peer\_tile\_wait} and \textit{peer\_tile\_notify} at lines 19 and 26, respectively. This example uses static mapping and demonstrates how to program communications in two directions: producer-consumer and peer-to-peer.

\begin{figure}[!t]
\centering
\includegraphics[width=0.47\textwidth]{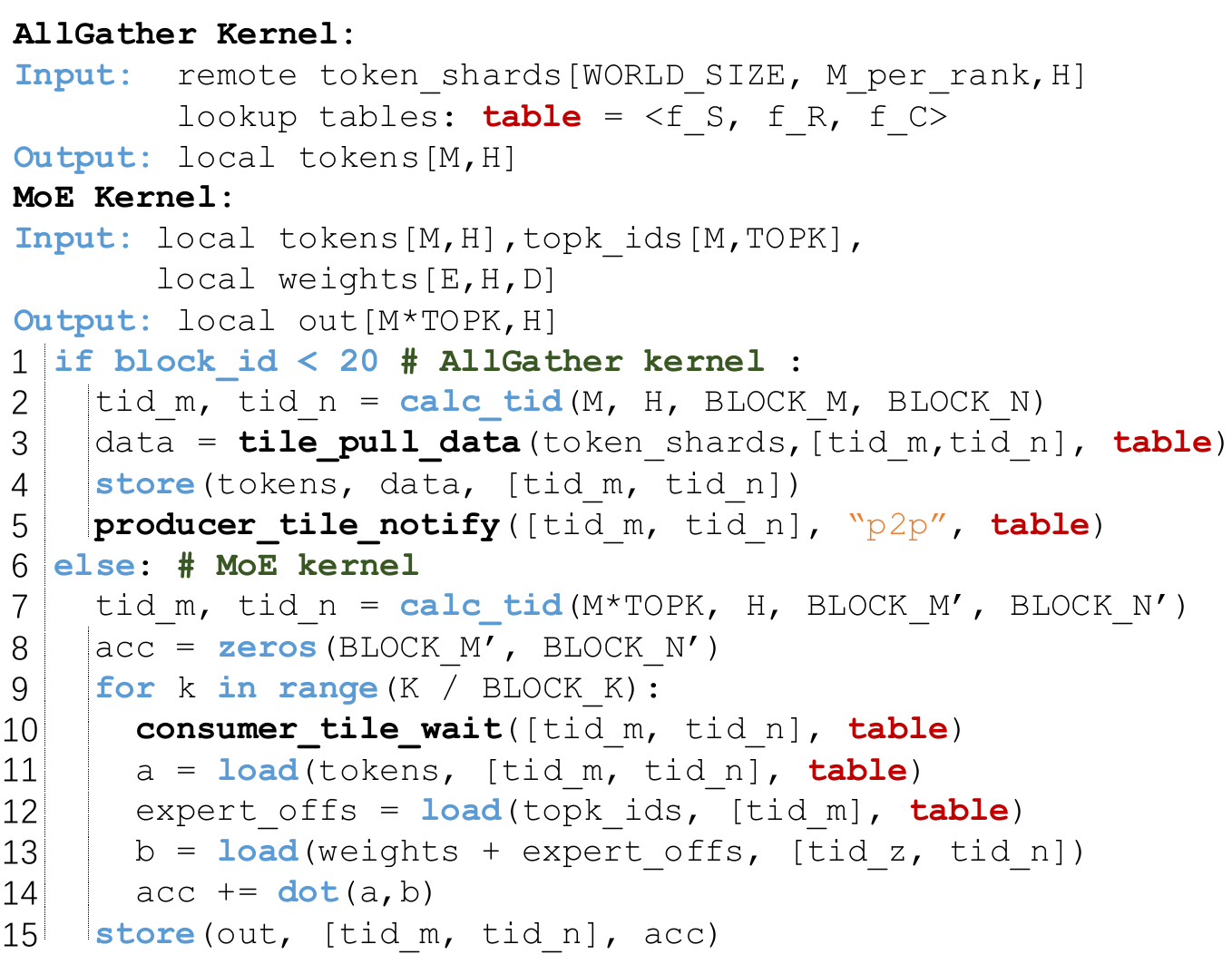}\vspace{-10pt}
\DeclareGraphicsExtensions.
\caption{AG + MoE overlapping kernel using \ours{}.}
\label{fig:ag_moe}
\end{figure}

Figure~\ref{fig:ag_moe} shows the pseudo code for AllGather + MoE. Again, both computation and communication use SMs, and we use 20 SMs for communication (see line 1). Note that MoE requires dynamic routing (\textit{topk\_ids} in inputs) to select experts for each token, necessitating dynamic mapping. We use \textit{table} to denote the lookup tables for shape mapping, rank mapping, and channel mapping. All the primitives involved should take \textit{table} as arguments so that \ours{} can generate correct code using the dynamic mappings. Additionally, the \textit{table} is required by the \textit{load} primitive because it uses the shape mapping in \textit{table} to gather the correct tokens (line 11) and the correct \textit{top\_ids} (line 12) for the corresponding tokens, which are needed by the current tile.

\begin{figure}[!t]
\centering
\includegraphics[width=0.48\textwidth]{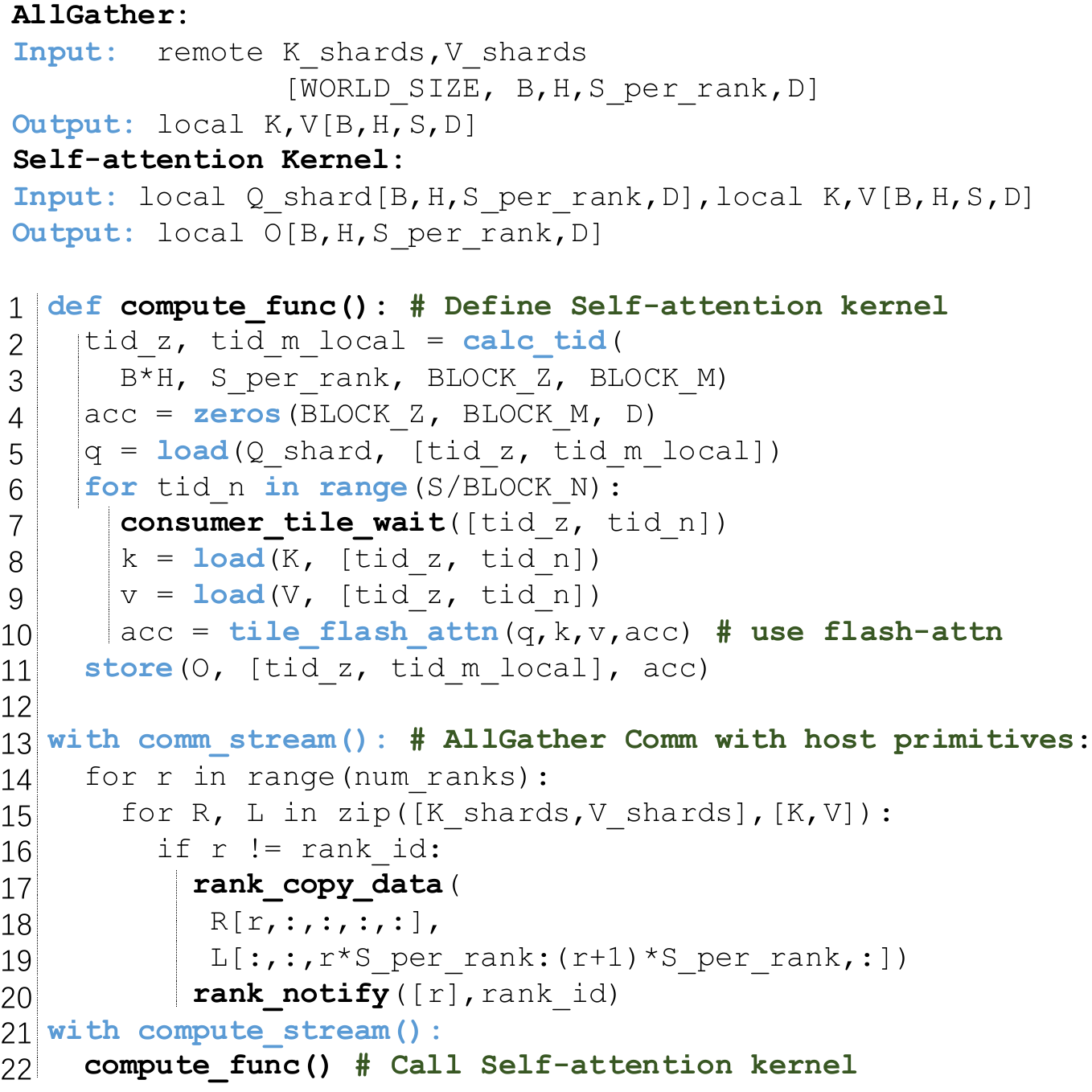}
\DeclareGraphicsExtensions.
\caption{AG KV + self-attention overlapping Kernel.}
\label{fig:sp-attn}
\end{figure}

Figure~\ref{fig:sp-attn} shows the pseduo code for AllGather KV + self-attention (sequence parallel). In this example, communication use copy engine. we use host primitives to trigger copy engines. The communication and computation run on two different streams. Communication is done using \textit{rank\_copy\_data} primitive, and the tile size for communication part is simply divide KVCache sequence length (\textit{S}) by the total number of ranks (\textit{WORLD\_SIZE}). For computation part, the tile size is different. Tile-centric mapping is used to guarantee the correct barrier operations between communication and computation parts.

These examples show that \ours{} is flexible in overlapping kernel design and reduces programming effort, thanks to our tile-centric primitives and mappings.

%% file: contents/compiler.tex
\section{Implementation}

\ours{} is implemented in Python on top of Triton~\cite{triton}. We extend Triton’s language features by implementing tile-centric primitives at the Python level, while the tile-centric mapping mechanism is realized through Python Abstract Syntax Tree (AST) transformations. The current implementation can be readily adapted to other compiler frameworks such as TVM~\cite{tvm} and MLIR~\cite{mlir}.

\begin{figure}[t]
\centering
\includegraphics[width=0.49\textwidth]{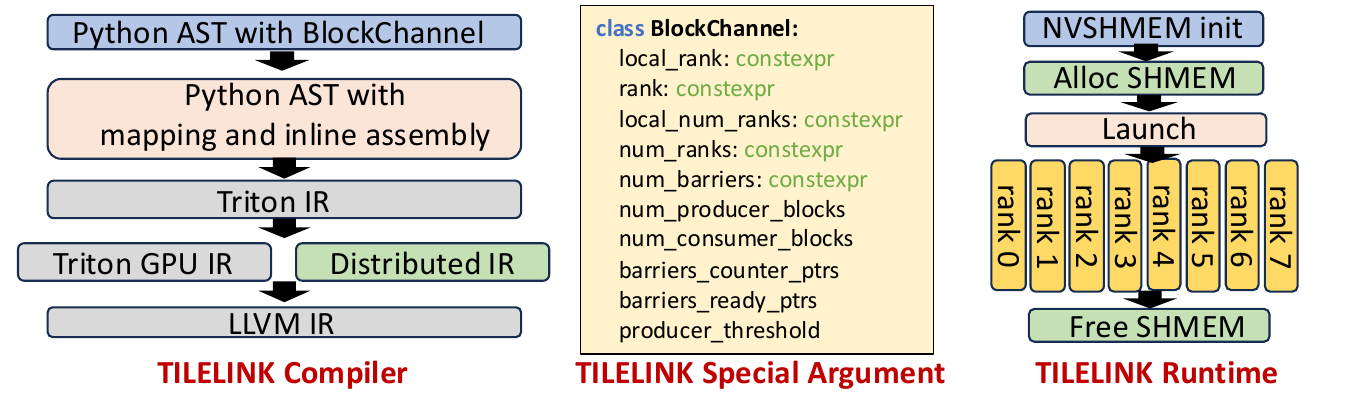}
\caption{Compilation and Runtime of \ours{}.}
\label{fig:compiler}
\end{figure}

As shown in Figure~\ref{fig:compiler}, the compiler takes as input a pure Python program combining \ours{}'s primitives with Triton's native primitives. A special parameter \textit{BlockChannel} is provided to serve as the tile-centric mapping context for computation and communication. 
The \textit{BlockChannel} parameter encapsulates distributed mapping metadata including current process rank, total world size, synchronization barrier configurations, and producer/consumer block relationships.
The Python program is parsed into an AST and translated into Triton IR. During translation, the \textit{BlockChannel} parameter is decomposed to construct the tile-centric mapping using embedded metadata. \ours{}'s primitives are converted into Triton's \textit{ElementwiseInlineAsmOp}. The Triton IR is then lowered to both Triton GPU IR and a new Distributed IR introduced by \ours{}. This Distributed IR is used to translate the special instructions expressed via \textit{ElementwiseInlineAsmOp} into LLVM IR, which is further compiled into PTX for NVIDIA GPUs. Support for additional backend architectures can be achieved by translating the LLVM IR into target-specific low-level assembly.
At runtime, NVSHMEM~\cite{nvshmem} is used to initialize the distributed execution environment and allocate shared memory. The generated code is launched across all processes to perform concurrent computation and communication, followed by proper shared memory deallocation after completion.

%% file: contents/evaluation.tex
\section{Evaluation}

\subsection{Experiment Setup}
In the evaluation, we use three benchmarks: MLP layer, MoE layer, and self-attention. The input shapes for these layers are listed in Table~\ref{table:config}. We use input configurations derived from real workloads such as LLaMA~\cite{llama3}, Gemma~\cite{gemma2}, and Qwen~\cite{qwen2}.
We use Async-TP PyTorch as the baseline for the decomposition method (Centauri~\cite{centauri} and Dist-Einsum~\cite{dist-enisum} are not publicly available), FLUX~\cite{flux} as the baseline for the fusion technique (CoCoNet~\cite{coconet} is available but its source code has been deprecated), and cuBLAS+NCCL as the baseline for non-overlap.
We use consistent parallel configurations for all the baselines.

\begin{table}[t]
  \centering
  \footnotesize
  \caption{Benchmark Shapes. \textit{S} is sequence length, \textit{H} is hidden dimension length, \textit{I} is intermediate size, \textit{E} is number of experts.}
\begin{small}
  \begin{tabular}{cccccc}
  \hline
    \multicolumn{6}{c}{{Configurations of MLP}} \\
  
  \hline
    \textbf{Name} & \textbf{S} & \textbf{H} & \textbf{I} & \multicolumn{2}{c}{\textbf{Source Model}} \\ 

  \hline
    MLP-1 & 8192 & 4096 & 11008 & \multicolumn{2}{c}{LLaMA-7B} \\
    \hline
    MLP-2 & 8192 & 4096 & 14336 & \multicolumn{2}{c}{LLaMA-3.1-8B} \\
    \hline
    MLP-3 & 8192 & 3584 & 14336 & \multicolumn{2}{c}{Gemma-2-9B} \\
    \hline
    MLP-4 & 8192 & 4608 & 36864 & \multicolumn{2}{c}{Gemma-2-27B} \\
    \hline
    MLP-5 & 8192 & 8192 & 28672 & \multicolumn{2}{c}{LLaMA-3.1-70B} \\
    \hline
    MLP-6 & 8192 & 8192 & 29568 & \multicolumn{2}{c}{Qwen-2-72B} \\

  \hline
    \multicolumn{6}{c}{{Configuration of MoE}} \\

  \hline
    \textbf{Name} & \textbf{S} & \textbf{H} & \textbf{I} & \textbf{E} & \textbf{topk}\\ 

  \hline
    MoE-1 & 8192 &	2048 &	1536 &	8 &	2\\

  \hline
    MoE-2 & 8192 &	2048 &	1536 &	32 &	2\\

  \hline
    MoE-3 & 8192 &	2048 &	1536 &	32 &	5\\

  \hline
    MoE-4 & 8192 &	4096 &	2048 &	8 &	2\\

  \hline
    MoE-5 & 8192 &	4096 &	2048 &	32 &	2\\

  \hline
    MoE-6 & 8192 &	4096 &	2048 &	32 &	5\\

  \hline
    \multicolumn{6}{c}{{Configuration of self-attention}} \\

  \hline
    \textbf{Name} & \textbf{heads} & \textbf{head dim}  & \multicolumn{3}{c}{\textbf{sequence length choices}} \\

  \hline
    Attn-1 &  32 & 128 &	\multicolumn{3}{c}{16k, 32k, 64k, 128k}	\\
  \hline
    Attn-2 & 64 &	128 & \multicolumn{3}{c}{16k, 32k, 64k, 128k}	\\
  \hline
  \end{tabular}
  \label{table:config}
\end{small}
\end{table}

\subsection{Single Layer Performance}

\begin{figure}[!t]
\centering
\includegraphics[width=0.49\textwidth]{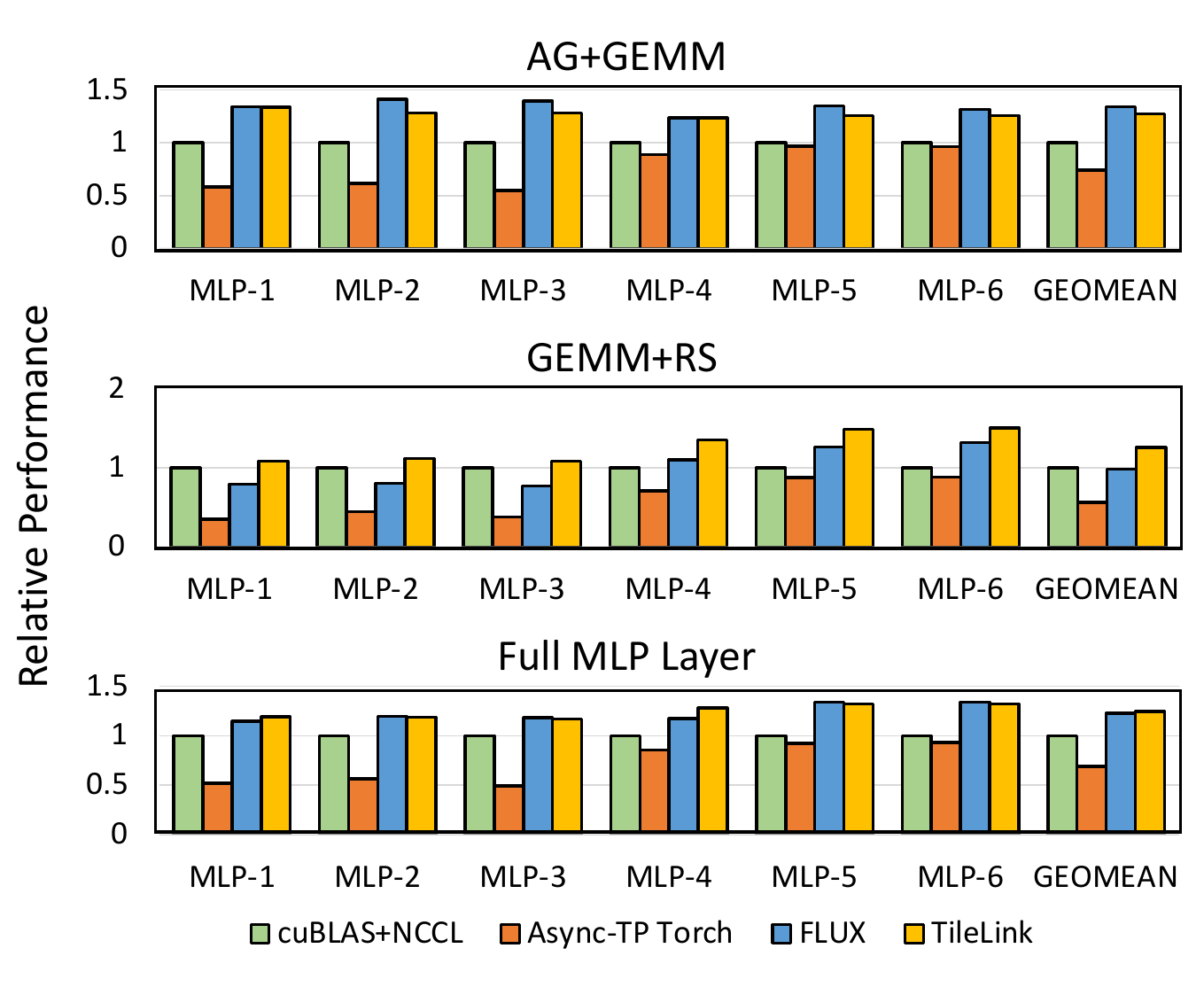}
\DeclareGraphicsExtensions.
\caption{Performance Results of MLP layers (AG+GEMM and GEMM+RS) on 8$\times$H800.
}
\label{fig:results-mlp}
\end{figure}

\textbf{MLP Layer:} the MLP layer is composed of two parts, the first part is mainly composed of AllGather + GEMM, the second part is mainly composed of GEMM + ReduceScatter, there is one activation layer (e.g., SiLUMul or GeLUMul) between these two parts. We evaluate the two parts separately and also evaluate the full performance of the MLP layer. The results on 8$\times$H800 cluster are shown in Figure~\ref{fig:results-mlp}. 
For AG + GEMM, Async-TP PyTorch cannot produce a speedup because the decomposed GEMMs are too small to fully utilize the device. Also, according to our tracing results, Async-TP PyTorch uses too many host-driven synchronizations and thus incurs non-negligible overhead to the overlapped kernel. FLUX achieves the highest speedup ($1.34\times$ over cuBLAS+NCCL) due to its highly optimized implementation. \ours{} also achieves a speedup over cuBLAS+NCCL ($1.27\times$), reaching $94.5\%$ of FLUX's performance. Note that \ours{} only requires hundreds of lines of Python code, while FLUX requires thousands of lines of CUDA code. The overlapped kernel generated by \ours{} maps AllGather to the DMA engine.

As for GEMM + ReduceScatter, \ours{} gives the best performance: $1.25\times$ over cuBLAS+NCCL, $2.22\times$ over Async-TP PyTorch, and $1.28\times$ over FLUX. \ours{} decouples the design space of GEMM and ReduceScatter, enabling each part to find their best optimizations, while FLUX uses a tightly coupled fusion kernel for this case, which performs sub-optimally in evaluation.
The overlapped kernel generated by \ours{} maps the ReduceScatter to both DMA engine and SMs (streaming multiprocessors), which is a hybrid resource mapping: scatter is done using DMA, and reduction is done on SMs.
Combining both parts with intermediate activation, \ours{} achieves performance comparable to FLUX (101.4\%) and a $1.24\times$ speedup over cuBLAS+NCCL. These results show that \ours{} can achieve performance comparable to state-of-the-art fusion libraries with significantly less code (as pointed out in the motivational example of this paper).

\begin{figure}[!t]
\centering
\includegraphics[width=0.49\textwidth]{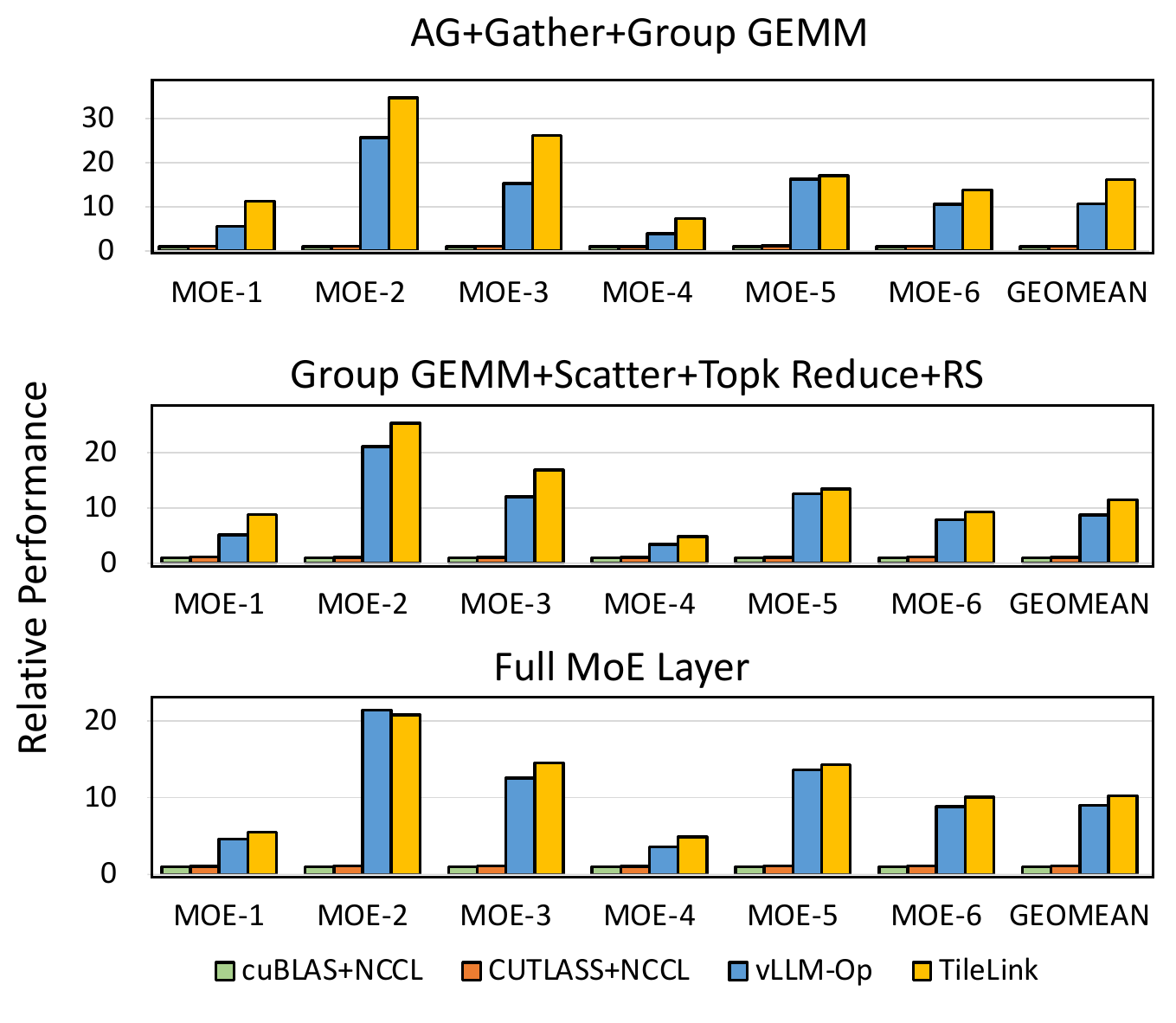}
\DeclareGraphicsExtensions.
\caption{Performance Results of MoE layers (AG + Gather + GroupGEMM and GroupGEMM + Scatter + Reduce + RS) on 8$\times$H800.
}
\label{fig:results-moe}
\end{figure}

\textbf{MoE Layer:} The MoE layer is much more complex than MLP layers and requires dynamic mapping during compilation. The MoE layer can also be divided into two parts: AG + Gather + Group GEMM and Group GEMM + Scatter + Topk Reduce + RS. There is a Gather operator in the first part and a Scatter + Topk Reduce operator in the second part because the dynamic routing shuffles tokens to different experts. These two operators can be fused into Group GEMM kernels. vLLM~\cite{vllm} provides implementations for such fused Group GEMM operations.

For the first part, Figure~\ref{fig:results-moe} shows the evaluation results. The cuBLAS and CUTLASS baseline implementations do not fuse the gather and scatter operations into Group GEMM, 
resulting in a performance 
bottleneck. The results from vLLM show that such fusion can improve performance by $9.82\times$. \ours{} achieves even better performance than vLLM (an average of $1.51\times$ improvement) because, in addition to the gather-scatter fusion, \ours{} also overlaps communication with computation. In the code generated by \ours{}, AllGather is mapped to the DMA engine.

\begin{figure}[!t]
\centering
\includegraphics[width=0.49\textwidth]{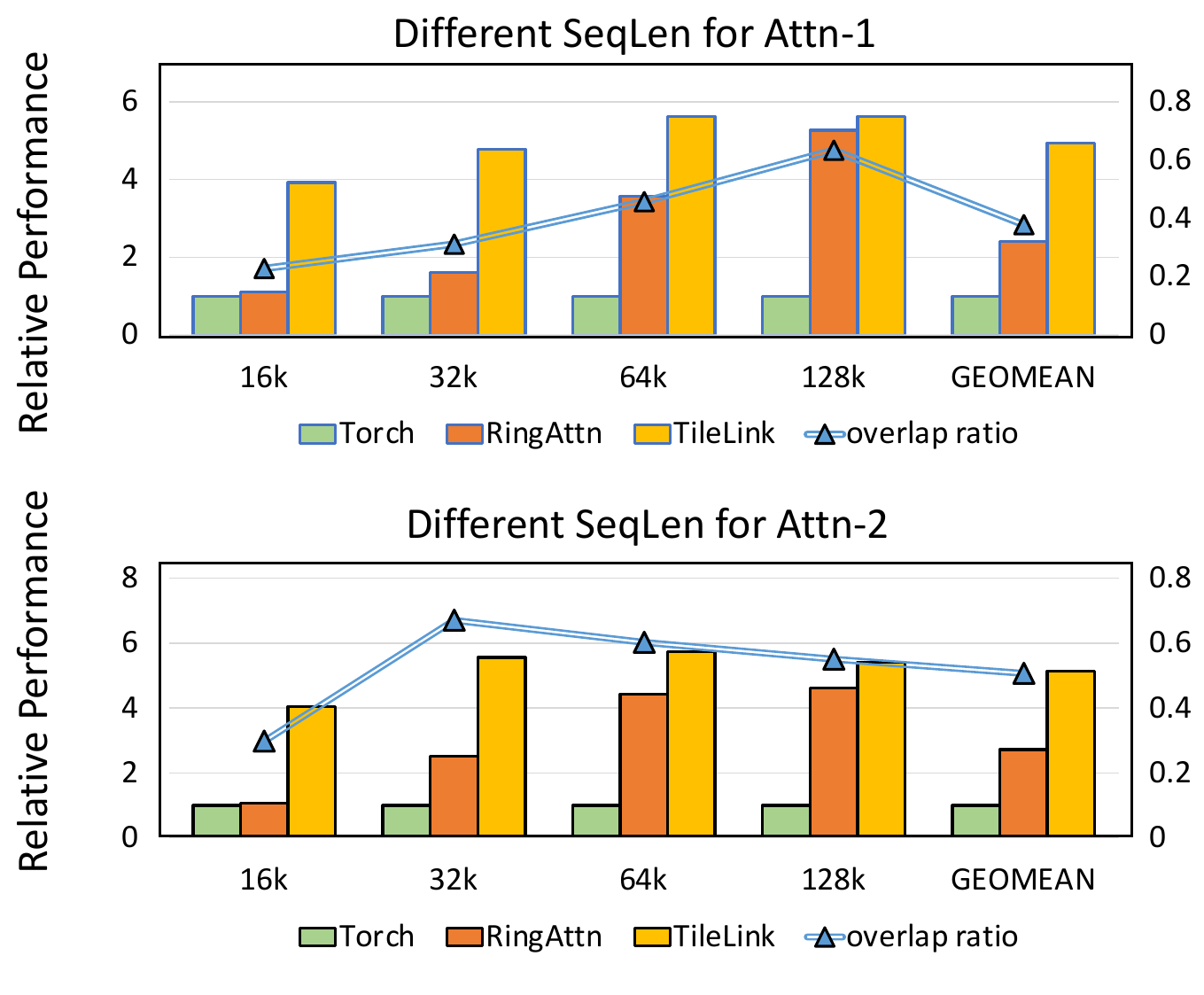}
\DeclareGraphicsExtensions.
\caption{Performance Results and Overlap Ratio of Self-attention Layers on 8$\times$H800.
}
\label{fig:results-attn}
\end{figure}

For the second part, \ours{} achieves an average speedup of $1.31\times$ over vLLM and $10.56\times$ speedup over CUTLASS+NCCL. This part of MoE has two epilogues: Topk Reduce and RS. \ours{} overlaps three kernels using the tile-centric primitives, demonstrating that the primitives are versatile enough to create extended producer-consumer chains in practice.
\ours{} maps Topk Reduce to SMs, and maps ReduceScatter to both the DMA engine and SMs.
The full MoE performance is also shown in Figure~\ref{fig:results-moe}, on average, \ours{} achieves a $1.14\times$ speedup over vLLM. The maximal speedup over cuBLAS+NCCL is $20.76\times$.
Note that existing libraries such as FLUX and Async-TP PyTorch do not support overlapping MoE layers. \ours{} supports MoE thanks to its flexible primitives and dynamic mappings.

\textbf{Self-Attention Layer:} Self-attention is composed of two batch GEMMs and one softmax, which are often fused together using Flash-Attention techniques~\cite{flashattn-2}. Sequence-parallel self-attention consists of an AllGather component and a self-attention computation component. We first implement Flash-Attention in \ours{} on Hopper GPUs and then use \ours{} primitives to overlap AllGather and Flash-Attention. The performance results are shown in Figure~\ref{fig:results-attn}. We test self-attention with different sequence lengths, from 16k to 128k, covering both short and long contexts. \ours{} shows consistent speedups over both the PyTorch non-overlap implementation (Torch) and RingAttention~\cite{ringattn} (RingAttn) across all the sequence lengths. On average, \ours{} achieves a $5.04\times$ speedup over Torch and a $1.97\times$ speedup over RingAttn.

We also plot the overlap ratio for self-attention, where overlap ratio is defined as
\begin{equation*}
\begin{small}
\begin{aligned}
ratio = \frac{comp\_only\_time + comm\_only\_time - overlap\_time}{comm\_only\_time}.\\
\end{aligned}
\end{small}
\end{equation*}
Overlap ratio can be used to measure how much communication overhead is hidden after overlapping. The results in Figure~\ref{fig:results-attn} shows that \ours{} can effectively overlap $43.9\%$ communication overhead on average.

\subsection{End-to-End Evaluation}

\begin{figure}[!t]
\centering
\includegraphics[width=0.49\textwidth]{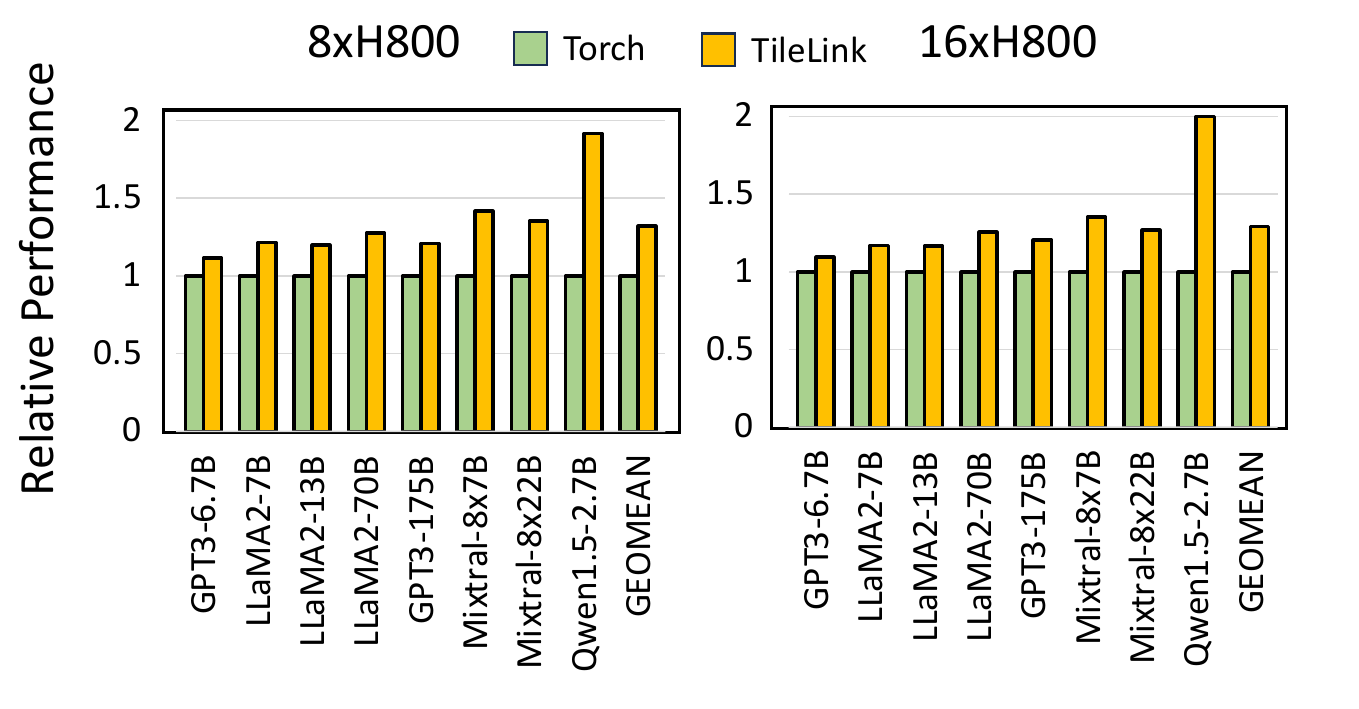}
\DeclareGraphicsExtensions.
\caption{Performance Results of End-to-end Models on 8$\times$H800 and 16$\times$H800.
}
\label{fig:results-e2e}
\end{figure}

We integrate \ours{} into PyTorch and evaluate end-to-end performance for 8 different LLMs on H800 clusters. We first evaluate the performance on a single node with 8$\times$H800 GPUs. The results are shown in the left part of Figure~\ref{fig:results-e2e}. The first five LLMs are dense models, while the other three models are MoE models. Qwen1.5 uses shared experts in MoE, we combine MLP layer and MoE layer together to support shared experts. We use batch size 4 and sequence length 8192. The results show that on average, \ours{} achieves a $1.32\times$ speedup over the PyTorch baselines. The average speedup of dense models is $1.20\times$, which aligns well with the speedup of single layer MLP. Although \ours{} achieves good speedups for self-attention, MLP layers dominate the performance of end-to-end evaluation (note that there are also large MLP layers before and after self-attention layer).
The average speedup of MoE models is $1.54\times$, which is lower than the speedup of a single MoE layer. In MoE models, MLP layers and MoE layers each occupy about $50\%$ of the total execution time, so the final speedup lies between the speedup of MLP and the MoE.

We also deploy \ours{} for multi-node evaluation. Tensor parallel is often used within one node due to the low inter-node bandwidth. So we use data parallel between two nodes and use tensor parallel in each node. The results on two nodes with 8$\times$H800 GPUs show similar outcomes to those on a single node, as expected. We double the batch size for this evaluation. The overall speedup is $1.29\times$, which is slightly lower than a single node due to additional communication overhead between two nodes.

\subsection{Discussion}
\textbf{Support for model-level communication:} \ours{} can be extended to support model-level parallelism (e.g., pipeline parallelism). To achieve this, we can integrate NVSHMEM functionalities into \ours{}'s \textit{tile\_push\_data} primitive and follow the same compilation techniques as \ours{}. We leave this for future work.

\textbf{Support multiple backends:} Currently, \ours{} targets only NVIDIA GPUs. To support more hardware, we can extend the low-level compilers (e.g., TVM, which supports more hardware than Triton), while keeping the primitives and compilation techniques of \ours{} unchanged.

%% file: contents/related-work.tex
\section{Related Work}

Compute-communication overlapping has been studied for years. Early work focuses on CPU clusters with MPI programming model~\cite{overlap-1, overlap-2, overlap-3, overlap-4}. With the fast advancement of LLMs, overlapping computation and communication on AI accelerators such as GPUs and NPUs has been proposed~\cite{coconet, dist-enisum, flux, amd-fused, centauri, pytorch}.

\textbf{Decomposition-based overlapping} focuses on splitting operators into smaller ones and overlapping them by rearranging asynchronous pipelines. Dist-Einsum~\cite{dist-enisum} implements overlapping kernels for MLP layers on Google TPUs; Async-TP PyTorch~\cite{torchtitan} provides implementations of overlapped AllGather GEMM and GEMM ReduceScatter; Centauri~\cite{centauri} systematically explore the three-level design space composed of model, layer, and operation overlapping. Decomposition-based method enables fast development and good compatibility with existing frameworks.

\textbf{Fusion-based overlapping} uses kernel fusion techniques to fuse computation kernel with communication kernel. CoCoNet~\cite{coconet} first proposes to fuse CUTLASS GEMM with NCCL kernels and produces state-of-the-art performance on V100 GPUs; FLUX~\cite{flux} follows the idea of CoCoNet and implements high-performance overlapped kernels on A100 and H800 GPUs; another fusion library from AMD~\cite{amd-fused} implements various overlapped kernels for DLRM and LLM on AMD GPUs. These studies require a long time to develop due to the lack of high-level programmable primitives. Compared to them, \ours{} provides flexible primitives and achieves comparable performance.

\textbf{Overlapping compilers} use compilation techniques to generate efficient overlapped kernels. CoCoNet compiles high-level operators and schedules into invocations of low-level CUTLASS GEMM and NCCL kernels. Dist-Einsum compiles DNN graphs to device code by decomposing original large operators into small operators and inserting synchronizations among them. These compilers provide little or no programming control for optimization choices such as tile sizes, tile orders, and resource bindings. On the other hand, code generation compilers~\cite{tvm, triton, mlir} and auto-tuners~\cite{flextensor, ansor, amos, tensorir} provide mature code generation support for single device. 
Recent work from AMD~\cite{amd-fused} also use Triton to generated overlapping kernels.
Pallas~\cite{pallas} is a distributed compiler that generates Triton code through compilation and supports computation-communication overlapping. However, the overlapping feature is currently only available on Google TPUs, not GPUs.
\ours{} provides a set of tile-centric primitives and automatically compiles them into device code using tile-centric mappings, supporting a wide range of workloads.

%% file: contents/conclusion.tex
\section{Conclusion}
To deploy large DNN models on distributed systems, overlapping communication and computation is of vital importance. Previous overlapping studies either bring sub-optimal performance or have difficulty in developing high-performance kernels. In this paper, we propose \ours{} to generate high-performance overlapped kernels. \ours{} uses a set of tile-centric primitives to enhance productivity and uses tile-centric mappings to generate low-level code. In experiments, \ours{} achieves from $1.17\times$  to $20.76\times$ speedups over non-overlapping baselines and comparable performance to state-of-the-art overlapping libraries.